\documentclass[preprint,12pt]{elsarticle}

\usepackage{latexsym,graphicx,epsf,float,afterpage,natbib}
\usepackage[T1]{fontenc}        % euro quality fonts [T1] (togeth. w/ textcomp)
\usepackage{textcomp, amssymb}  % additional symbols (there are more packages)
\usepackage{setspace}           % doublespacing
\usepackage{anysize}            % margin package sets tighter margins
\usepackage[all]{xy}            % creating figures within latex
\usepackage[tight]{subfigure}% subfigures: figures within figures
\usepackage{makeidx}                       % for \printindex
\makeindex                                 % creates paper.idx index file
\usepackage{tikz}
\usetikzlibrary{shapes,arrows}
\usepackage{enumerate}

\def\AssX#1{{\setbox0=\hbox{#1\sf\textbf{A}}\lower.2\ht0\copy0}} % helper for \Ass
\usepackage{graphicx}
\usepackage{amsmath}
\usepackage{amssymb}
\usepackage{dsfont}
\usepackage{bbold}
\usepackage{mathrsfs}
\usepackage{leftidx}
\usepackage{framed}
\usepackage{filecontents}
\usepackage{caption}

\usepackage[english]{babel}

\usepackage{psfrag}
\usepackage[subrefformat=parens,labelformat=parens,singlelinecheck=false,font=normalsize]{subfig}
\usepackage{xcolor}

\journal{...}

\makeatletter
\def\@author#1{\g@addto@macro\elsauthors{\normalsize%
    \def\baselinestretch{1}%
    \upshape\authorsep#1\unskip\textsuperscript{%
      \ifx\@fnmark\@empty\else\unskip\sep\@fnmark\let\sep=,\fi
      \ifx\@corref\@empty\else\unskip\sep\@corref\let\sep=,\fi
      }%
    \def\authorsep{\unskip,\space}%
    \global\let\@fnmark\@empty
    \global\let\@corref\@empty  %% Added
    \global\let\sep\@empty}%
    \@eadauthor={#1}
}
\makeatother

\begin{document}

\begin{frontmatter}



  \title{Lumped parameter modelling of ferroelectric ceramics for control applications using simulink}

%\author{Author One\corref{cor1}\fnref{fn1}}
%\ead{email@uni.edu}
%\fntext[fn1]{Student}

\author{S. Maniprakash}
\ead{maniprakash.subramanian@udo.edu}

%\author{R. Jayendiran$^c$}
%\author{A. Menzel$^{a,b}$}
%\author{A. Arockiarajan$^c$}

%\fntext[label2]{ppo}

\address{Institute of Mechanics, TU Dortmund, Germany}
%\address{$^b$Division of Solid Mechanics, Lund University, Sweden}
%\address{$^c$Department of Applied Mechanics, IIT Madras, India}

\begin{abstract}

Simplifying the constitutive behaviour of a material in terms of the lumped parameter elements is useful to design plant models in control engineering. In this contribution, a lumped parameter modelling approach is used to represent the constitutive behaviour of ferroelectric ceramics. Using the elements available in simulink, an electrical circuit is designed to simulate the ferroelectric behaviour. The simulation results of dielectric and butterfly hysteresis shows the possibility of applying the lumped parameter modelling approach in the design of displacement control systems.

%Further development and design of piezoelectric composites enhances
%the improved use of piezoelectric materials and devices by overcoming
%their brittleness. In order to engineer this class of materials and to
%predictably simulate its behaviour, a computationally efficient
%constitutive model is established in this work. This contribution
%deals with the development of a model for piezoelectric composites to
%capture their effective behaviour.
% We first discuss a
%three-dimensional fully coupled electromechanical rate-dependent
%model for the response of ferroelectric ceramics. Secondly, a simple homogenisation approach
%%an equivalent \textcolor{red}{layered approach}
% is applied to capture the behaviour of
%composites for various volume fractions of PZT fibres under different loading
%frequencies. Following this, a finite element formulation is applied
%in order to study the behaviour of composites. Finally, these two
%approaches are compared with experimental results.

\end{abstract}

\begin{keyword}
%\sep 
  ferroelectrics \sep switching \sep lumped parameter model \sep coupled
  problems \sep control engineering


\end{keyword}

\end{frontmatter}


%\input{ijss1.tex}

%\chapter*{1-3 ferroelectric composites}\label{chapter:1-3 ferroelectric composites} %%%%%%%%%%%%%%%%%%%%%%%%%%%%

\section{Introduction}
Due to its strong electromechanical coupling nature, the piezoelectric material can be used as sensors and actuators. Vibration control, structural health monitoring, energy harvesting are few application regimes of this material to mention \cite{C3EE42454E,smith2005a,C3EE42454d}. Above a certain threshold limit of electrical loading, the material shows a nonlinear hysteretic behaviour, which is known as ferroelectric behaviour. Obtaining a plant model for the ferroelectric behaviour is important for the design of control system applications based on large actuation displacements. In this work, from the view point of plant design for ferroelectric material in closed loop control systems, a lumped parameter model for ferroelectric ceramics is developed to capture the dielectric and butterfly hysteresis behaviour.

\section{Lumped parameter elements}
\begin{figure}[H]
\centering
\begin{subfigure}{}
\includegraphics[width=40mm]{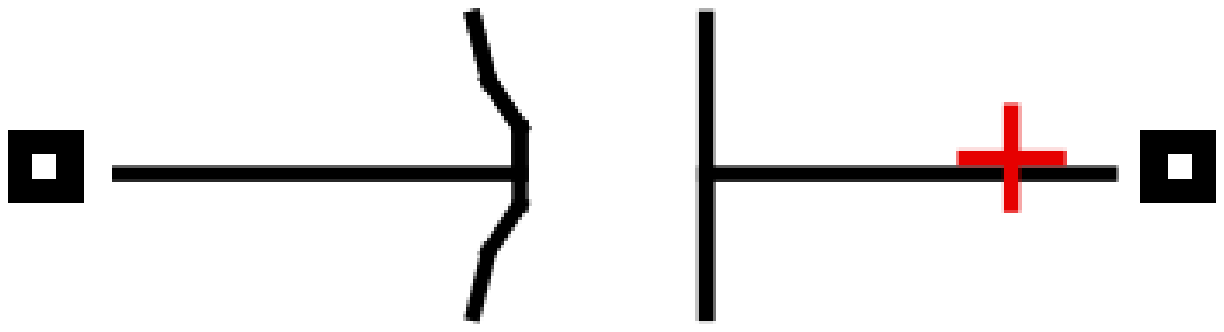}
\end{subfigure}
\begin{subfigure}{}
\includegraphics[width=40mm]{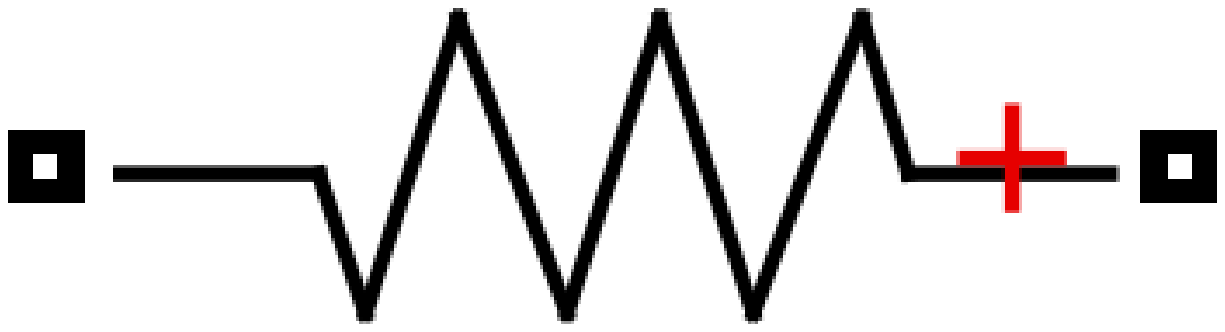}
\end{subfigure}
\begin{subfigure}{}
\includegraphics[width=40mm]{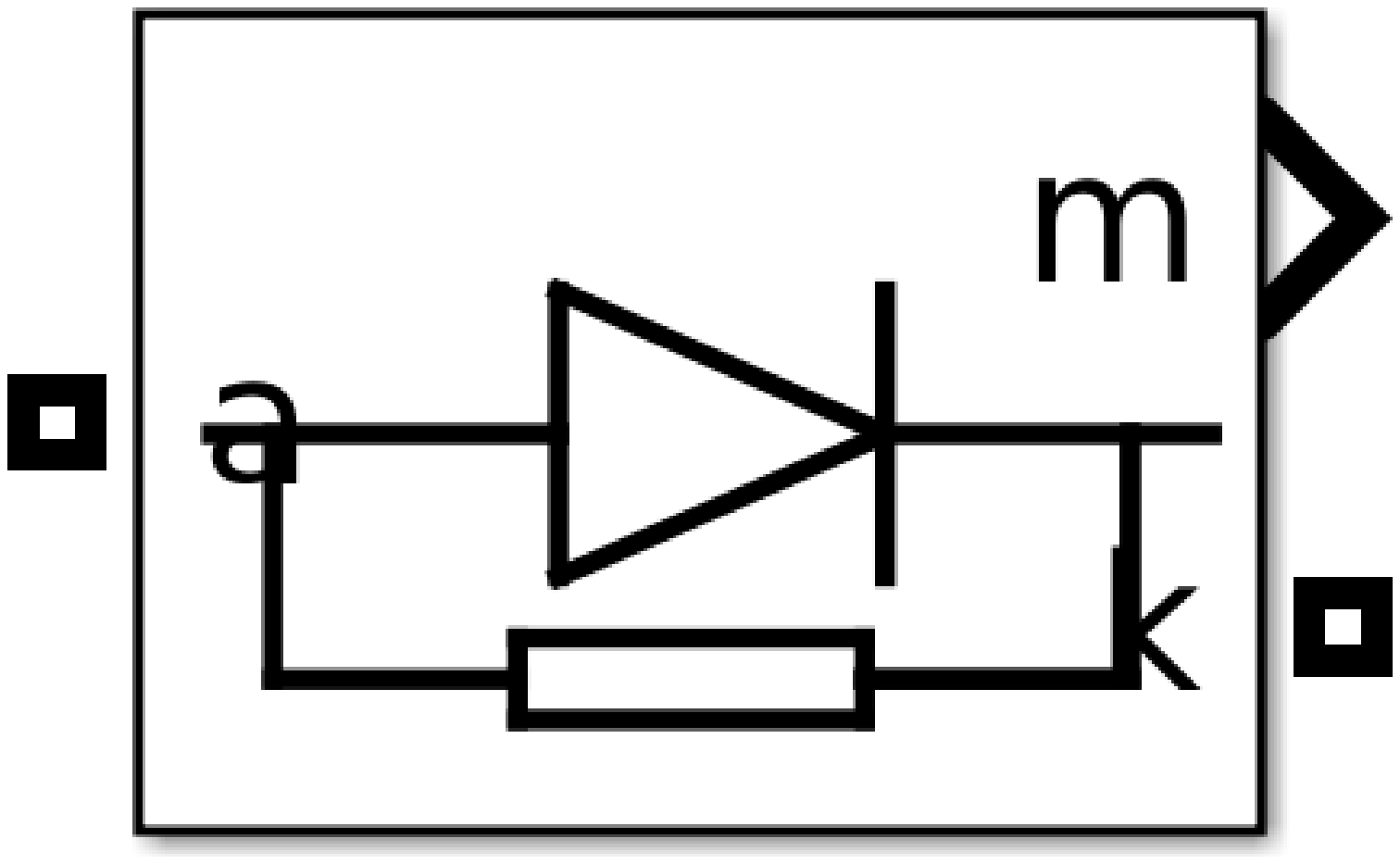}
\end{subfigure}
\\
\footnotesize{\hfill (a) Capacitor \hfill (b) Resistor \hfill (c) Diode \hfill}
\newline
\caption{Lumped parameter elements used for ferroelectric models}
\label{fig:lump_elem}
\end{figure}

In this lumped parameter modelling approach for ferroelectrics, the model is represented by a closed loop electrical circuit. The input signal is given by the applied voltage, $\phi$. The electrical output is given as the total charge supplied, $q$, to the circuit. The mechanical output is represented as strain. Fig.~\ref{fig:lump_elem} shows the different types of lumped parameter elements used for developing a constitutive model of ferroelectric ceramics.

\subsection{Capacitor}
The element capacitor represents the electrical energy stored in the material. The capacitance, $C$, of this element expresses the constitutive parameter of the material. The constitutive relation of this component can be written as
\begin{equation}
q = C\, \phi
\end{equation}

\subsection{Resistor}
The resistor element in the model addresses the dissipation of electrical energy in the system. Ohm's law represents the constitutive relation of this material as
\begin{equation}
I(t) = \dfrac{\partial q}{\partial t} = \dfrac{1}{R}\, \phi
\end{equation}
where $I$ is the electrical current across the element, $R$ represents the electrical resistance and $t$ represents the time variable.

\subsection{Diode}
Ferroelectric switching is the main reason for the nonlinear hysteresis behaviour. This switching phenomena occurs only if the applied voltage exceeds a certain threshold limit. 
To mimic this switching effect in the material, the diode element is introduced, which functions as a forward biased diode in the electrical circuit. The constitutive function of this element is given as
\begin{equation}
    \text{circuit} \Rightarrow
\begin{cases}
    \text{close},& \text{if } \phi \geq \phi_c\\
    \text{open},              & \text{otherwise}
\end{cases}
\end{equation}
where the scalar value $\phi_c$ represents the threshold limit voltage for switching.

\section{Constitutive circuits}
In this section, the constitutive relation of the ferroelectric material is presented using the above introduced lumped parameter elements. First, the constitutive circuit for piezoelectric material is discussed. Later, the extension of piezoelectric constitutive circuit to ferroelectric constitutive circuit is presented.

\subsection{Piezoelectric constitutive circuit}
\begin{figure}
 \centering
\includegraphics[width=120mm]{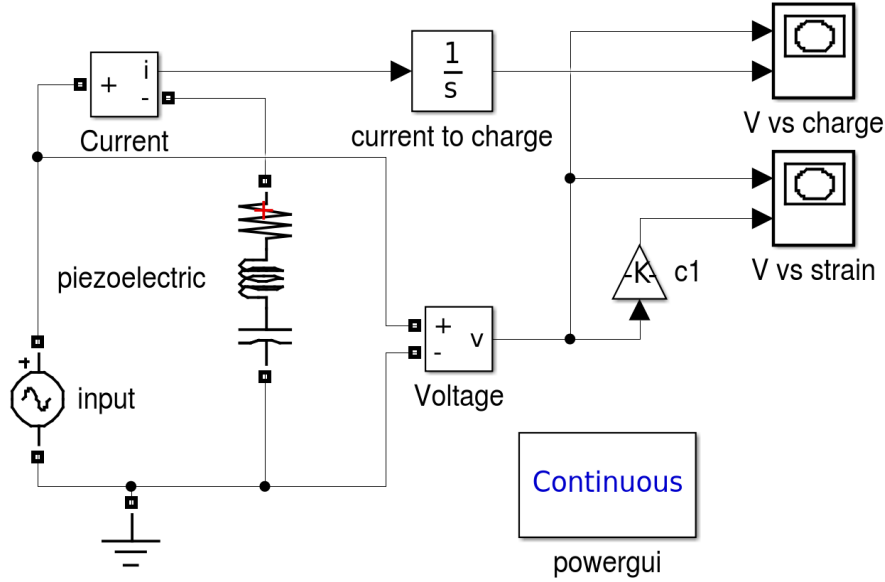}
\caption{Piezoelectric constitutive circuit.}
\label{fig:piezo_circuit}
\end{figure}

In the absence of external stress, the piezoelectric constitutive equations can be written as 

\begin{align}
q = C\, \phi\,,
\\
\varepsilon = c_1 \, \phi\,,
\end{align}
where $\varepsilon$ represents the strain value and the coupling factor $c_1$ represents the ratio of exerted strain to the applied voltage. This constitutive relation is obtained by constructing the electrical circuit as shown in Fig.~\ref{fig:piezo_circuit}. In this circuit, a series RLC load is connected. The value of resistance and inductance in this series can be chosen to be reasonably small so that the element can behave as a capacitor of lumped parameter elements. Since the strain is proportional to the applied voltage in the piezoelectric material, the strain value is calculated by applying a gain factor 'c1' in the circuit.

\subsection{Ferroelectric constitutive circuit}
\begin{figure}
\includegraphics[width=1.35\textwidth, angle =90 ]{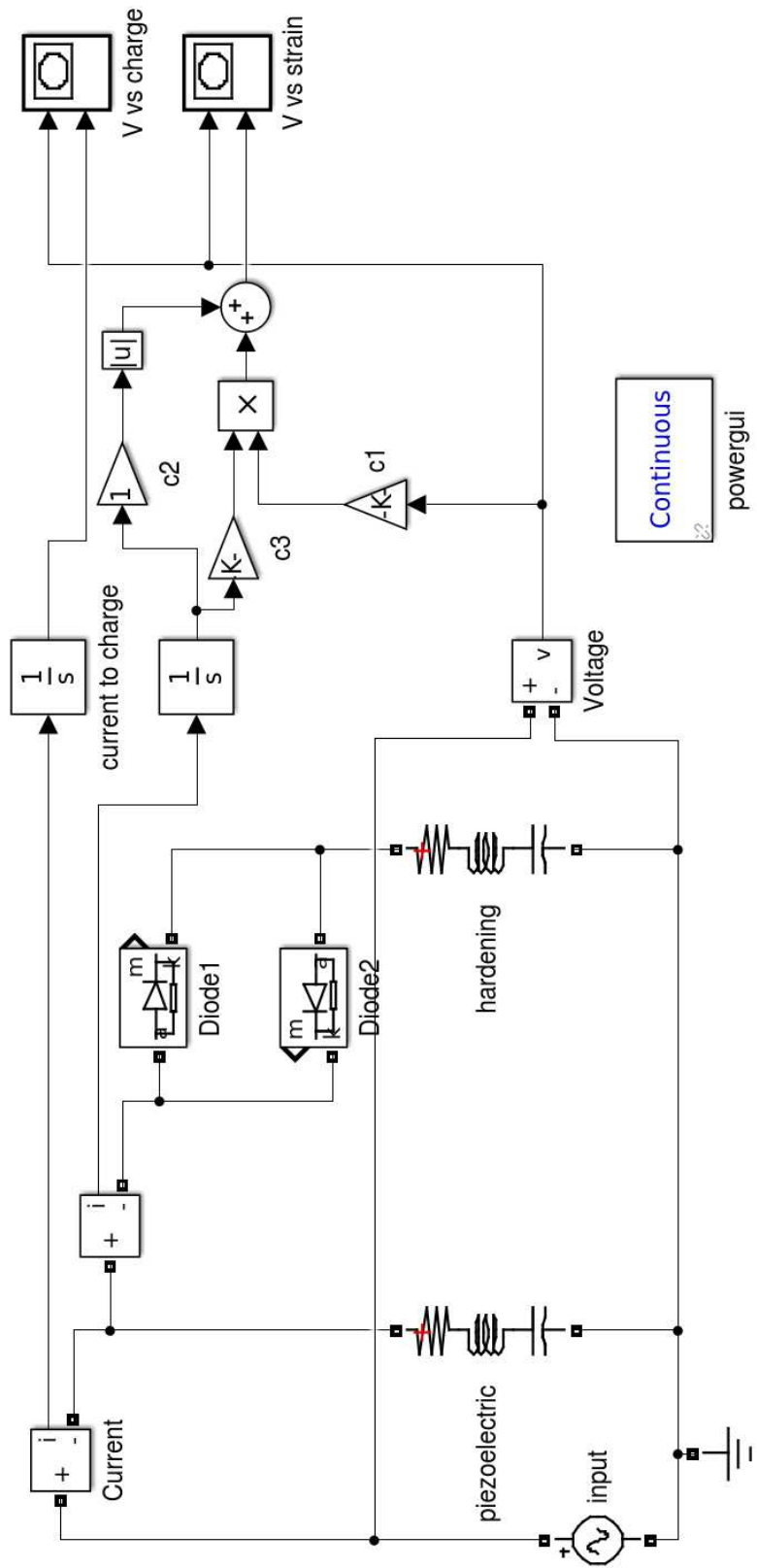}
\caption{Ferroelectric constitutive circuit.}
\label{fig:ferro_circuit}
\end{figure}

To impose and obtain the ferroelectric behaviour in this electrical circuit, a new circuit branch parallel to the piezoelectric RLC branch is introduced as shown in Fig.~\ref{fig:ferro_circuit}. In this branch, a series RLC load and two parallel diodes are connected in series. The diodes function as a switching condition of ferroelectric constitutive relation. The series RLC load acts as a hardening function. Negligible value of resistance of this RLC load can be chosen to model the rate-independent ferroelectric behaviour. To calculate remnant strain of the material, a one to one relation between remnant strain and remnant polarisation is used. This one to one relation is obtained in the circuit by multiplying the remnant polarisation with the gain factor 'c2' and subsequently by taking the modulus. Total strain can be obtained by adding the remnant strain and the strain obtained from piezoelectric coupling term. However, the value of the piezoelectric coupling coefficient $C$ depends on the value of remnant polarisation. Therefore to obtain the varying coupling coefficient, a new branch with the gain factor 'c3' is introduced to normalise the remnant polarisation.

\section{Results and conclusion}

\begin{figure}[H]
\centering
\begin{subfigure}{}
\includegraphics[width=70mm]{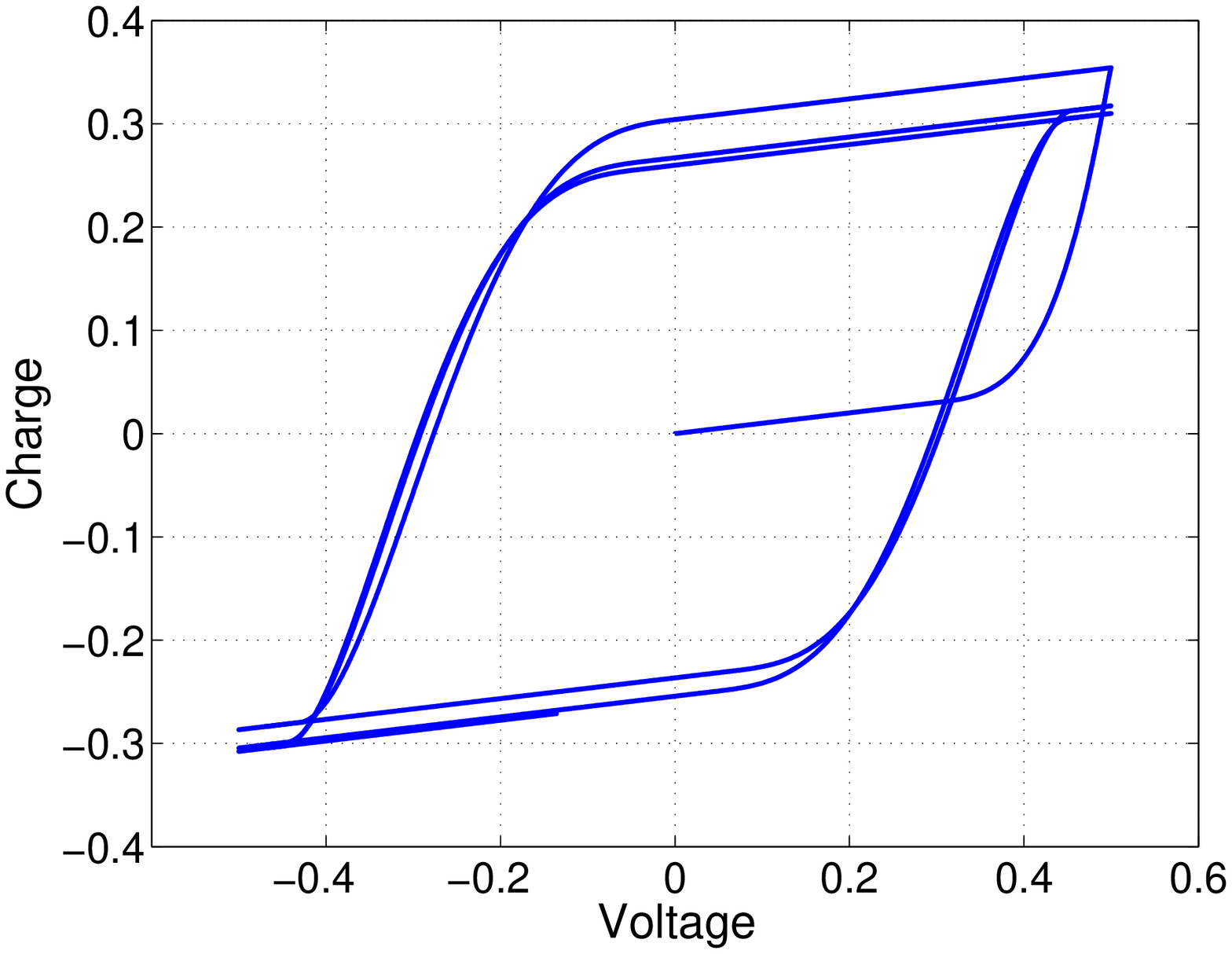}
\end{subfigure}
\begin{subfigure}{}
\includegraphics[width=70mm]{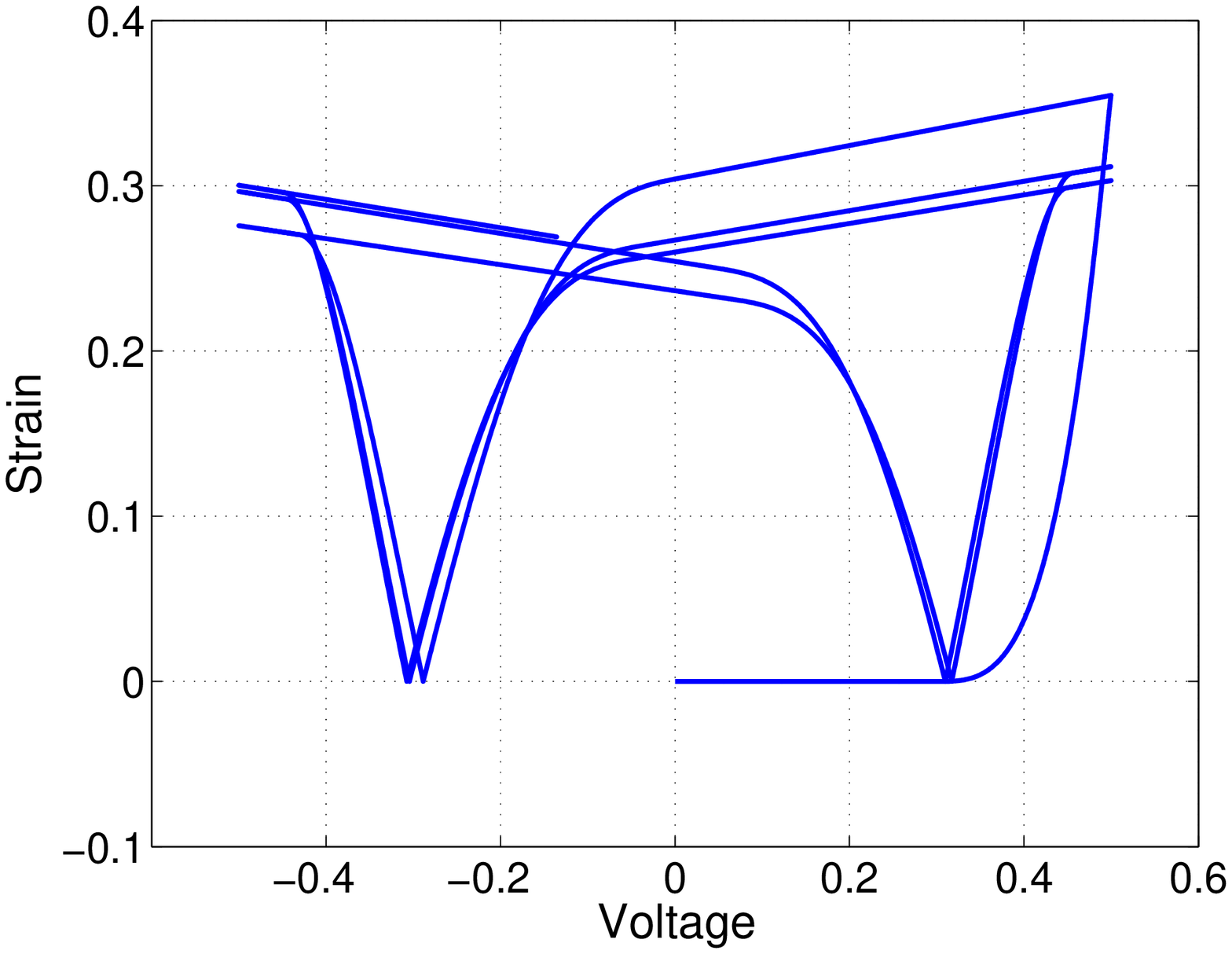}
\end{subfigure}
\\
\footnotesize{\hfill Dielectric Hysteresis \hfill \hfill Butterfly hysteresis \hfill}
\newline
\caption{Simulation results of ferroelectric lumped parameter model.}
\label{fig:results}
\end{figure}

Fig.~\ref{fig:results} shows the simulation results of the ferroelectric constitutive circuit obtained for the sinusoidal voltage input. The obtained results show that the discussed lumped parameter model can capture the dielectric hysteresis as well as butterfly hysteresis very well. Therefore, from the results, one could see the robustness in capturing the ferroelectric constitutive behaviour by using the simple lumped parameter modelling approach. This modelling approach will also be useful in addressing the other phenomena of ferroelectric materials and also offers deep insight and new ideas on the phenomenological constitutive model construction. Moreover to that, such simple models will be very useful in designing the control systems of actuator applications.

%\input{constitutive.tex}
%\input{composite_result.tex}
%\input{references.tex}
%\newpage

%\subsection*{Acknowledgement}
%
%Financial support by the NRW Graduate School of Energy Efficient
%Production and Logistics is gratefully acknowledged.
%%
%Partial financial support was also provided by the German Research Foundation(DFG) within the research unit 1509 "Ferroic Functional Materials" under project P6, which is gratefully acknowledged.

%\section*{References}
%\label{References}



%  \bibliographystyle{elsarticle-harv} 
%  \bibliography{ijssref}


%\begin{thebibliography}{00}


%\input{biblio.tex}

\bibliographystyle{plainnat}
\bibliography{references.bib}
%\bibliographystyle{elsart-num}
%\bibliography{\jobname}

%\end{thebibliography}
%\newpage
%\input{fig.tex}

%\tableofcontents

\end{document}